\documentclass[twocolumn,showpacs,superscriptaddress,amsmath,amssymb]{revtex4}
\usepackage{graphicx}
\usepackage{epsfig}
\usepackage{dcolumn}
\usepackage{bm}
\usepackage{color}
\usepackage{subfigure}

\def \jp {J/\psi}

\def \ff {\phi\phi}
\def \oo {\omega\omega}
\def \of {\omega\phi}

\def \gg {\gamma\gamma}
\def \chisq {\chi^2_{4C}}
\def \kk {K^+K^-}
\def \pp {\pi^+\pi^-}
\def \ppp {\pp\pi^0}
\newcommand{\chicj}{\chi_{cJ}}
\newcommand{\chiz}{\chi_{c0}}
\newcommand{\chio}{\chi_{c1}}
\newcommand{\chit}{\chi_{c2}}

\newcommand{\psip}{\psi(3686)}

\begin{document}

\title{\boldmath Observation of $\chio$ decays into vector meson pairs $\ff$, $\oo$, and $\of$}

\author{
{\small
M.~Ablikim$^{1}$, M.~N.~Achasov$^{5}$, L.~An$^{9}$, Q.~An$^{36}$, Z.~H.~An$^{1}$, J.~Z.~Bai$^{1}$, R.~Baldini$^{17}$, Y.~Ban$^{23}$, J.~Becker$^{2}$, N.~Berger$^{1}$
% ...
\\
\vspace{0.2cm}
(BESIII Collaboration)\\
\vspace{0.2cm}
{\it
$^{1}$ Institute of High Energy Physics, Beijing 100049, P. R. China\\
$^{2}$ Bochum Ruhr-University, 44780 Bochum, Germany\\
% ...
$^{5}$ G.I. Budker Institute of Nuclear Physics SB RAS (BINP), Novosibirsk 630090, Russia\\
% ...
$^{9}$ Guangxi University, Naning 530004, P. R. China\\
% ...
$^{17}$ Laboratori Nazionali di Frascati - INFN, 00044 Frascati, Italy\\
% ...
$^{23}$ Peking University, Beijing 100871, P. R. China\\
% ...
$^{36}$ University of Science and Technology of China, Hefei 230026, P. R. China\\
% ...
}} \vspace{0.4cm} }

\begin{abstract}

Using $(106\pm4)\times 10^6$ $\psip$ events accumulated with the
BESIII detector at the BEPCII $e^+e^-$ collider, we present the first measurement of decays of $\chio$ to vector meson pairs $\ff$, $\oo$, and $\of$.  The branching fractions are measured to be $(4.4\pm
0.3\pm 0.5)\times 10^{-4}$, $(6.0\pm 0.3\pm 0.7)\times
10^{-4}$, and $(2.2\pm 0.6\pm 0.2)\times 10^{-5}$, for
$\chio\to \ff$, $\oo$, and $\of$, respectively, which indicates that the hadron helicity selection rule is
significantly violated in $\chicj$ decays.  In addition, the measurement of $\chicj\to \of$
provides the first indication of the rate of doubly OZI-suppressed $\chicj$ decay.  Finally, we present improved measurements for the branching fractions of $\chi_{c0}$ and
$\chi_{c2}$ to vector meson pairs.

\end{abstract}


\maketitle

Decays of the $\chicj~(J=0,1,2)$ P-wave charmonium states are considered  to be an ideal laboratory to test
QCD theory. The initial theoretical calculations of $\chicj$ exclusive decays into
light hadrons predicted branching fractions that were smaller than the
experimental measurements~\cite{earlier}. With the inclusion of
the color-octet mechanism~\cite{octet}, calculations of
$\chicj$ decays into pairs of pseudoscalar mesons and pairs of baryons came into reasonable agreement with the experimental measurements, indicating the
importance of the color-octet mechanism.

In the case of $\chicj$ decays into pairs of vector ($J^{PC}=1^{--}$) mesons $VV$, where $V$ is an $\omega$ or $\phi$, the branching fractions for $\chi_{c0/2}$ decays to $\ff$ and $\oo$ have been measured to be at
the $10^{-3}$ level~\cite{besiiff,besiioo}, which is much larger
than predictions based on perturbative QCD calculations~\cite{pingrg}.
Decays of the $\chio$ into $ \ff$, $\oo$ and $\of$ violate the helicity selection rule (HSR) and are expected to be highly
suppressed~\cite{hsr}.
In addition, the decays $\chicj\to \of$ are doubly OZI
suppressed and have yet to be observed. Recently, long-distance effects in $\chi_{c1}$ decays \cite{zhaoqiang,chendy}
have been proposed to account for the HSR violation. Precise measurements of $\chi_{c1}\to VV $ decays will
help clarify the influence of long distance effects in this energy region.

In this Letter, we report measurements of $\chicj$ decays into
$\ff$, $\oo$, and $\of$ modes, where $\phi$ is reconstructed from $\kk$ or $\ppp$,
$\omega$ from $\ppp$, and $\pi^0$ from $\gg$. The data samples used in this  analysis
 consist of  $(106\pm4) \times 10^6$ $\psi^{\prime}$ decays and  42.6~pb$^{-1}$ of continuum data at
 $\sqrt s= 3.65$~GeV acquired with the BESIII detector~\cite{bes3}.
 The cylindrical core of the BESIII detector consists of a
helium-gas-based Main Drift Chamber~(MDC), a plastic scintillator
Time-of-Flight system~(TOF), a CsI(Tl) Electromagnetic
Calorimeter~(EMC), and a muon counter. The
charged particle and photon acceptance is $93\%$ of $4\pi$, and
the charged particle momentum and photon energy resolutions at
1~GeV are $0.5\%$ and $2.5\%$, respectively. The BESIII detector is modeled with a Monte Carlo (MC) simulation based on {\sc geant4}\cite{geant4a,geant4b}. The optimization of the event selection and the estimation of physics backgrounds are
performed with Monte Carlo simulations of $\psip$ inclusive/exclusive decays \cite{psipdecays}.

The final states of interest are $\gamma 2(\kk)$, $5\gamma
2(\pp)$, and $3\gamma\kk\pp$. Event candidates are required to
have four well reconstructed charged tracks with net charge zero, and at least
one, five, or three good photons, for $\ff$, $\oo$, and $\of$,
respectively.

Electromagnetic showers in BESIII detector are reconstructed from clusters of energy deposits in the EMC.
The energy deposited in nearby TOF counters is included to improve the reconstruction efficiency.
 A good photon is a shower in the barrel region
($|\cos\theta|<0.8$) with at least 25~MeV energy deposition, or in
the endcaps ($0.86<|\cos\theta|<0.92$) with at least 50~MeV
energy deposition, where $\theta$ is the polar angle of the shower. Showers in the region between the barrel and
the endcaps are poorly measured and excluded. Timing
requirements are used in the EMC to suppress electronic noise and energy deposits
unrelated to the event.

Charged tracks are reconstructed from MDC
hits. Each charged track is required to be in the polar angle
region $|\cos\theta|<0.93$ and to pass within $\pm 10$~cm of the
interaction point in the beam direction and within $\pm 1$~cm in
the plane perpendicular to the beam.

A  kinematic fit constrained by the initial $e^+e^-$ four-momentum in the lab frame is applied to the decay hypotheses
$\psip\to\gamma 2(\kk)$, $5\gamma2(\pp)$, and $3\gamma\kk\pp$. The final state photons are identified with the photon-charged-track combination
 that has a minimum $\chi^2_{4C}$ value (for definition of $\chi^2_{4C}$, see \cite{yellowbook}) when sampling all candidate photons. The vertex of all charged tracks must be consistent with the measured beam interaction point. The $\chi^2_{4C}$ selection efficiency is optimized using the ratio of signal to backgrounds  in the data: $\chisq<60$ for
$\gamma 2(\kk)$, $3\gamma\kk\pp$,  and $\chisq<200$ for $5\gamma2(\pp)$
is required. To separate the
$K^{\pm}$ from $\pi^{\pm}$ in the $3\gamma\kk\pp$ final state, two
kaons are identified with the requirements that $P(K)>P(\pi)$ and $P(K)>P(p)$, where
$P(X)$ is the probability of hypothesis $X$ as evaluated from the TOF and $dE/dx$ information.

The mass windows for resonance candidates are set according to the optimized ratio of signals to backgrounds in the data. The $\pi^0$ candidates are selected by requiring $0.1 < M_{\gamma\gamma}
< 0.15$~GeV/$c^2$.  The $\phi$ and $\omega$ candidates are selected by requiring
$|M_{\kk}-1.019|<0.015$~GeV/$c^2$,
$|M_{\pp\pi^0}-1.019|<0.030$~GeV/$c^2$, and
$|M_{\pp\pi^0}-0.783|<0.050$~GeV/$c^2$, for $\phi\to \kk$,
$\phi\to \ppp$, and $\omega\to \ppp$, respectively.

\begin{figure}
  \centering
 \includegraphics[width=0.45\textwidth,height=11cm]{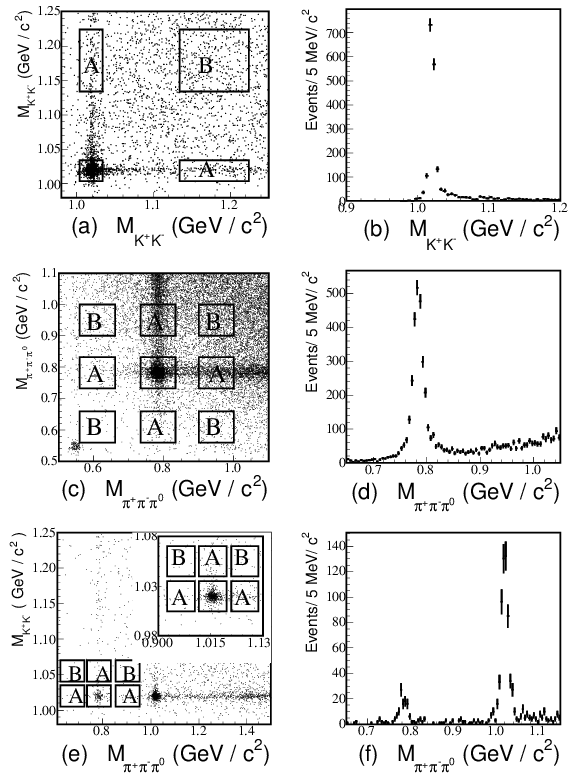}
\caption{\label{evtslct}The left column shows scatterplots for events within the $\chicj$ mass region. The boxes indicate
the signal region (without label) and sideband regions labeled as \textbf{A} and \textbf{B}. The plots in the right
column are the one-dimensional projections of the system recoiling against a selected $\phi$
or $\omega$ resonance. Plots (a) and (b) are for the $\gamma 2(\kk)$ mode; (c)
and (d) for the $5\gamma2(\pp)$ mode; and (e) and (f) for
the $3\gamma\kk\pp$ mode.}
\end{figure}

\begin{figure}
  \centering
 \includegraphics[width=0.42\textwidth,height=9cm]{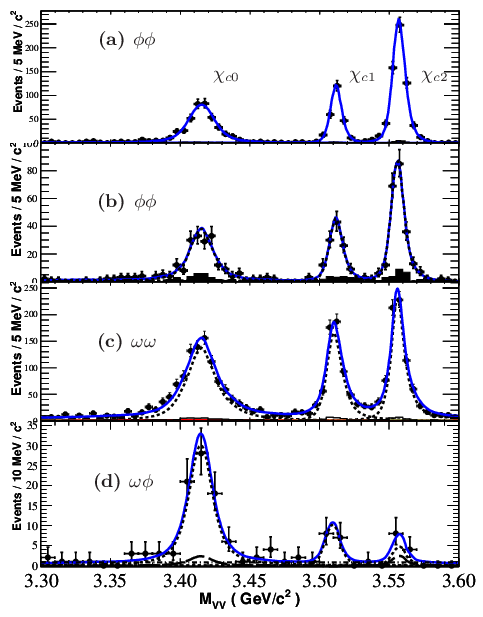}
\caption{Invariant mass of $VV$ for (a) $\ff$ mode in the $\gamma2(\kk)$ final state, (b) $\ff$ mode in the $\gamma\ppp\kk$ final state,  (c) $\oo$ mode in the $\gamma2(\ppp)$ final state, and (d) $\of$ mode in the $\gamma\ppp\kk$ final state.  The points with error bars are the data; the solid lines are the fit results; and dotted lines represent the signal components.  The shaded and open histograms  in (a,b) and (c), respectively, are peaking backgrounds.  In (c), the shaded histogram denotes the non-$\chicj$ backgrounds.  In (d) the long dash line is background normalized by a simultaneous fit to $\of$ sidebands, and the dash-dot line is non-$\chicj$ background.\label{mchicj}}
\end{figure}

For $\chicj\to \ff\to 2(\kk)$, the two $\phi$ candidates with the
minimum value of $(M_{\kk}^{(1)}-1.019)^2+(M_{\kk}^{(2)}-1.019)^2$ are
taken as the signal. No artificial $\phi$-pair peaks are produced when this selection criteria is applied to MC simulation of the process $\chicj\to2(\kk)$.
A scatterplot of masses for one $\kk$ pair {\it versus} the other
$\kk$ pair is shown in Fig.~\ref{evtslct}(a), where a clear
$\phi\phi$ signal can be seen. The $M_{\kk}$ distribution, after
requiring that the other two kaons are consistent with being a $\phi$, is shown in
Fig.~\ref{evtslct}(b). A $\phi$ peak is clearly seen with
very low background. The $\ff$ invariant mass distribution for the selected events is shown in
Fig.~\ref{mchicj} (a), where $\chicj$ signals are
clearly observed. The MC simulation shows that the peaking backgrounds, {\it i.e.,} backgrounds that produce $\chicj$ signal peaks, are mostly from $\chicj\to\phi\kk$ and $2(\kk)$ final states; the backgrounds from misidentified charged particles are negligible.  The levels of the peaking
backgrounds are evaluated from $N_{AB}=r_A N_{A}^{dt}-r_B N_B^{dt}$, where $N_A^{dt}(N_B^{dt})$  is the number of data events falling into box \textbf{A} (\textbf{B}), as
indicated in Fig.~\ref{evtslct}(a), and the normalizing factors $r_i=N_{sig}^{MC}/N_i^{MC}$ with $i=A$ or $B$ are determined from MC simulation for modes $\chicj\to\phi\kk$ and $2(\kk)$, respectively. Here $N_{sig}^{MC}(N_{i}^{MC})$ is the number of MC events falling into the signal box (\textbf{A} or \textbf{B}).  These backgrounds will be indistinguishable from signal events; therefore, we fix their normalization, independently for each $\chicj$ peak, in the final fit.

To study $\chicj\to \oo$ decays into the
$2(\pp\pi^0)$ final state, two $\pi^0$ candidates are selected by minimizing the value of
$(M_{\gg}^{(1)}-0.135)^2+(M_{\gg}^{(2)}-0.135)^2$ when
 sampling all four-photon combinations from the selected five photons.
The $\ppp$ combination closest to the nominal $\omega$ mass is
taken as one $\omega$ candidate, and the remaining three pions are
assumed to be from the other $\omega$. No artificial $\omega$-pair peaks are produced from the application of this $\omega$-selection criteria to a MC simulation for $\chicj\to 2(\ppp)$. A scatterplot of the mass for one $\ppp$ pair {\it versus} the other $\ppp$
pair is shown in Fig.~\ref{evtslct}(c), and the $M_{\ppp}$
distribution for the three pions recoiling against an $\omega$ candidate is plotted
in Fig.~\ref{evtslct}(d). The $\oo$ mass
spectrum is shown in Fig.~\ref{mchicj} (c), where
$\chicj$ signals are prominent.  The MC simulation shows that the backgrounds in the $\oo$ signal region include peaking backgrounds from $\chicj\to\omega\ppp$ and $2(\ppp)$, and non-peaking backgrounds from the $\psip$ decays into the same final states without intermediate $\chicj$ states. The backgrounds from misidentified charged particles are negligible. Potential backgrounds from $\chicj\to\ff\to 2(\ppp)$ and $\chi_{c0/2}\to\eta\eta\to 2(\ppp)$ do not survive our selection criteria.
As in the $\chicj\to\ff$ mode, the sizes of the peaking backgrounds from $\chicj\to\omega\ppp$ and $2(\ppp)$ are evaluated by selecting data events located in sideband boxes \textbf{A} and \textbf{B}, respectively, as indicated in Fig.~\ref{evtslct}(c). The peaking backgrounds are normalized according to the ratio of MC events falling into the signal region and those falling into the sidebands. The normalization of these peaking backgrounds is fixed in the final fit.

To study $\chicj\to \of$ and $\ff$ decays into the
$\kk\ppp$ final state, the photon pair with invariant mass closest to the $\pi^0$
nominal mass is taken as the $\pi^0$ candidate. A scatterplot
of masses for $\kk$ pairs {\it versus} that for $\ppp$ pairs is shown in
Fig.~\ref{evtslct}(e), and the $M_{\ppp}$ distribution for events satisfying
 $\phi\to \kk$  is shown in Fig.~\ref{evtslct}(f),
where the $\omega\to \ppp$ and $\phi\to \ppp$ signals are clearly
seen. The $\ff$ and $\of$  mass spectra are shown in
Figs.~\ref{mchicj} (b) and \ref{mchicj} (d),
respectively. Similar to the case for $\chicj\to\ff\to2(\kk)$, the peaking backgrounds from the $\chicj\to\phi\ppp$ or $\phi\kk$, and $\kk\ppp$ are evaluated by selecting data events falling into
sideband boxes \textbf{A} and \textbf{B}, respectively,  as indicated in the inserted plot in
Fig.~\ref{evtslct}(e). The peaking backgrounds are normalized according to the ratio of MC events falling into the signal region and those falling into the sidebands. The normalization of these peaking backgrounds is fixed in the final fit.

 The numbers of observed events are obtained by fitting the $M_{VV}$ distributions. The observed line shapes are described with modified $\chicj$ MC shapes plus backgrounds. Possible interference effects between the signal mode and the peaking background modes are not considered for all modes. The original $\chicj$ MC shapes are generated by a relativistic Breit-Wigner incorporated with full helicity amplitudes in the EvtGen package \cite{evtgen}, and their masses and widths are set to the nominal values \cite{pdg2010}. In the fits they are modified by convolving them with Gaussian functions $G(M_{VV}-\delta M_J, \sigma_J)$, where $\delta M_J$ and $\sigma_J$ correct the $\chi_{cJ}$ mass and width or resolution, respectively, in the simulation. The values of $\delta M_J$ and $\sigma_J$, determined from the fits, are less than 1 MeV for all modes and from 1 to 5 MeV, respectively. Backgrounds from QED processes, which are estimated from the application of a similar analysis to the continuum data, are negligible. For $\chicj\to\phi\phi$,  the peaking backgrounds are fixed to the sideband estimates as mentioned above, and other combinatorial backgrounds are parameterized by a second-order polynomial with parameters that are allowed to float in the fit.  For all modes, a maximum-likelihood technique\cite{roofit} is employed to estimate parameters. After projecting the best fit into the binned histograms shown in Fig. \ref{mchicj}, we determine $\chi^2/ndf=0.46$ for $\chicj\to\ff\to2(\kk)$ and $0.50$ for the $\chicj\to\ff\to\kk\ppp$, where $ndf$ is the number of degrees of freedom. The fitted results are plotted in Fig.~\ref{mchicj}(a) and (b), respectively. The numbers of signal events are listed in Table~\ref{brsummary}.

For the $\chicj\to\omega\omega$ channel, backgrounds include the peaking backgrounds estimated from $\omega$ sidebands indicated in Fig.~\ref{evtslct} (c), non-$\chicj$ backgrounds ($\psip\to\gamma\oo$) fixed at the normalized MC shape of phase space using the data information, and smooth combinatorial backgrounds that are parametrized by a second-order polynomial. The $\chi^2/ndf$ for the fit is 0.97. The fit results are shown in Fig.~\ref{mchicj} (c).

 To extract the signal yield, as well as to estimate the statistical significance for the $\chicj\to \of$ mode, a
simultaneous fit is performed to $M_{\of}$ distributions both in
$\of$ signal and sideband regions of boxes \textbf{A} and \textbf{B} [see Fig.~\ref{evtslct} (e)]. The peaking backgrounds are normalized according to the ratio of MC events falling into the signal region to those falling into the sideband regions for the $\psip\to\gamma\phi\ppp,~\gamma\omega\kk$ and $\psip\to\gamma\kk\ppp$ events that are within the $\chicj$ mass region. Because of the low signal yield in this mode, the parameters $\delta M_J$ and $\sigma_J$ of the modified MC shapes are fixed at the values determined in the fit of $\chicj\to \ff\to \kk\ppp$. The $\chi^2/ndf$ is 0.62. The fit results are shown in
Fig.~\ref{mchicj} (d), and the numbers of signal events are listed in Table~\ref{brsummary}.

\begin{table}[htbp]
\caption{Summary of the branching fractions ($\mathcal B$)
for $\chicj\to\ff$, $\oo$, and $\of$. Here $N_\mathrm{net}$ is the number
of signal events, $\epsilon$ is the detection efficiency. The upper limit is estimated
at the 90\% C.L.\label{brsummary}}
\begin{center}
\begin{tabular}{llll}
 \hline\hline
 Mode&$N_\mathrm{net}$ & $\epsilon$~(\%) & $\mathcal B(\times10^{-4}$)  \\\hline
$\chiz\to\ff$&$433\pm23$&$22.4$&$7.8\pm0.4\pm0.8$\\
$\chio\to\ff$&$254\pm17$&$26.4$&$4.1\pm0.3\pm0.4$\\
$\chit\to\ff$&$630\pm26$&$26.1$&$10.7\pm0.4\pm1.1$\\
$\to2(\kk)$&&&\\\hline
$\chiz\to\ff$&$179\pm16$&$12.8$&$9.2\pm0.7\pm1.0$\\
$\chio\to\ff$&$112\pm12$&$15.3$&$5.0\pm0.5\pm0.6$\\
$\chit\to\ff$&$219\pm16$&$14.9$&$10.7\pm0.7\pm1.2$\\
$\to\kk\ppp$&&&\\\hline
Combined: &&&\\
$\chiz\to\ff$&---&---&$8.0\pm0.3\pm0.8$\\
$\chio\to\ff$&---&---&$4.4\pm0.3\pm0.5$\\
$\chit\to\ff$&---&---&$10.7\pm0.3\pm1.2$\\\hline
$\chiz\to\oo$&$991\pm38$&$13.1$&$9.5\pm0.3\pm1.1$\\
$\chio\to\oo$&$597\pm29$&$13.2$&$6.0\pm0.3\pm0.7$\\
$\chit\to\oo$&$762\pm31$&$11.9$&$8.9\pm0.3\pm1.1$\\
$\to2(\ppp)$&&&\\\hline
$\chiz\to\of$&$76\pm11$&$14.7$&$1.2\pm0.1\pm0.2$\\
$\chio\to\of$&$15\pm4$&$16.2$&$0.22\pm0.06\pm0.02$\\
$\chit\to\of$&$<13$&$15.7$&$<0.2$ \\
$\to\kk\ppp$&&&\\
\hline
 \hline
\end{tabular}
\end{center}
\end{table}

The uncertainties due to the modified $\chicj$ MC shapes are estimated  by replacing them with Breit-Wigner functions convolved with the instrumental resolution functions in the fits.  The quality of the resulting fit is not as good as using the modified MC shapes. The difference of signal yields varies from 1\% to 4\%, and this is included as a systematic error.

The detection efficiencies are determined from MC simulations for
the sequential decays $\psip\to \gamma\chicj\to VV$, $V$ decays
into the selected final state. The decays $\psip\to \gamma\chicj$
are generated by assuming a pure $E1$ transition.  The $\chicj\to
VV$ decays and subsequent decays of the $V$ are modeled with
helicity amplitudes that provide angular distributions consistent
with the data.

The systematic uncertainties on the $\chicj$ decay branching
fractions arise from the $\pi^{\pm}$ and $K^{\pm}$ tracking,
$K^{\pm}$ identification, EMC shower reconstruction, number of
$\psip$ decays, kinematic fitting, modified MC shapes, background estimation, $\chicj$
signal extraction and uncertainties from branching fractions of
$\psip\to \gamma\chicj$, $\phi\to \kk$, $\omega\to \pp\pi^0$ and
$\pi^0\to \gamma\gamma$.
The uncertainties caused by MDC tracking
are estimated to be  2\% for each charged track \cite{syserrors}. The
uncertainty due to $K^\pm$ identification is evaluated to be 2\% per kaon \cite{syserrors}.
The uncertainty due to the photon reconstruction is determined
 to be 1\% for each photon \cite{syserrors}.
 The uncertainty in the number of
$\psip$ decays is 4\%~\cite{psipdecays}. The uncertainties due to
the kinematic fit are determined by comparing the efficiency at
the given $\chi^2_{4C}$ values for the MC sample to control
samples selected from data, i.e, $\psip\to\gamma\ff\to\gamma2(\kk)$, $\psip\to\pi^0\pi^0\jp, ~\jp\to2(\pp),\pi^02(\pp)$ and  $\psip\to\pp\jp,~\jp\to\kk\pi^0$. The kinematic-fit uncertainty varies from 0.5\%
($\gamma2(\pp\pi^0)$ mode) to 3.7\% ($\gamma\kk\pp\pi^0$ mode).
The uncertainties of the peaking backgrounds for $\chicj\to\ff\to 2(\kk)$ are evaluated by
 comparing the sideband estimates to the exclusive MC simulation on the modes $\chicj\to\phi\kk$ and $2(\kk)$, while for other modes the uncertainties
 are estimated by varying the size of sideband boxes. The uncertainties of the peaking background estimates are less than 3\%. The uncertainty from the MC normalization factor is found to be negligibly small.
 The total systematic uncertainties are 10\% for
$\chicj\to\ff \to2(\kk)$ mode, and 11\% for $\chicj\to
\oo\to 2(\pp\pi^0)$, $\chicj\to \ff,\;\of\to \kk\pp\pi^0$ modes.

The branching fractions for $\chicj$ decays are determined from
$\mathcal B=N_\mathrm{net}/(N_{\psip}\epsilon \prod_i \mathcal B_i)$, where $N_\mathrm{net}$ and
$\epsilon$ are the number of net signal events and
the detection efficiency, respectively.  The detection efficiencies are listed in Table~\ref{brsummary}. Here $N_{\psip}=(106\pm
4)\times 10^{6}$~\cite{psipdecays} is the number of $\psip$
events,  and $\prod_i \mathcal B_i$ is the product of world average branching fractions values~\cite{pdg2010} for $\psip\to \gamma\chicj$ and the other
meson decays that are involved.  For the $\chicj\to\phi\phi\to\kk\ppp$ branching fraction we double the efficiency listed in Table~\ref{brsummary} since our analysis sums over the two combinations for each $\phi$ to decay to either $\kk$ or $\ppp$. The resulting branching fractions are
listed in Table~\ref{brsummary}. The statistical significance of $\chicj\to\of$ is derived from the change of $-2\ln \mathcal{L}$ obtained from fits with and without each of the three $\chicj\to\of$ signal components.  We obtain a significance of $4.1\sigma$ for $\chi_{c1}\to\of$ and $1.5\sigma$ for $\chi_{c2}\to\of$.  The significance of the $\chi_{c0}\to\of$ signal is $10\sigma$. Using the Bayesian method, the upper limit for the number of signal events of the $\chi_{c2}\to\of$ mode is 13 at the 90\% confidence
level (C.L.). The branching fractions for $\chicj\to \ff$ measured
in $2(\kk)$ and $(\kk)(\pp\pi^0)$ final states are
combined into a weighted average, where
common systematic uncertainties are counted only once.

In summary, the HSR suppressed decays of $\chi_{c1}\to\ff,~\oo$, and the doubly OZI suppressed decay $\chi_{c0}\to\of$ are observed for the first time.  The branching fractions are measured to be $(4.4\pm
0.3\pm 0.5)\times 10^{-4}$, $(6.0\pm 0.3\pm 0.7)\times
10^{-4}$, and $(1.2\pm 0.1\pm 0.2)\times 10^{-4}$, for
$\chio\to \ff$, $\oo$, and $\chi_{c0}\to\of$, respectively, We also find evidence for $\chi_{c1}\to \of$ decay with a signal significance of $4.1\sigma$.  The branching fractions for $\chi_{c0/2}\to\ff,\oo$ decays are remeasured with a precision that is better than those of the current world average values \cite{pdg2010}.  These precise measurements will be helpful for understanding $\chicj$ decay mechanisms. In particular, the measured branching fractions for $\chi_{c1}\to VV$ indicate that HSR is significantly violated and that long distance effects play an important role in this energy region. The long distance effects from the intermediate charmed meson loops in $\chi_{c1}\to \ff$ and $\oo$ decays \cite{zhaoqiang,chendy} can contribute to the branching fractions at the level of $10^{-4}$ but are more than an order of magnitude too small to explain the doubly OZI suppressed decay rate for $\chi_{c1}\to\of$ that we measure \cite{chendy}.

We thank the accelerator group and computer staff of
IHEP for their effort in producing beams and processing
data. We are grateful for support from our institutes and
universities and from these agencies: Ministry of Science
and Technology of China, National Natural Science
Foundation of China, Chinese Academy of Sciences,
Istituto Nazionale di Fisica Nucleare, Russian Foundation
for Basic Research, Russian Academy of Science (Siberian
branch), U.S. Department of Energy, and National
Research Foundation of Korea.


\begin{thebibliography}{99}

\bibitem{earlier} A. Duncan, A. Mueller, Phys. Lett. B {\bf93}, 119  (1980);  H. F. Jones,  J. Wyndham, Nucl. Phys. B {\bf 195}, 222 (1982); M. Anselmino, F. Murgia, Phys. Rev. D {\bf47}, 3977 (1993).
\bibitem{octet} J.~Bolz, P.~Kroll and G.~A.~Schuler,
Eur. Phys. J. C {\bf 2}, 705 (1998); S.~M.~H.~Wong,  Eur. Phys. J.
C {\bf 14}, 643 (2000).

\bibitem{besiiff} M.~Ablikim {\it et al.} (BES Collaboration),
Phys. Lett. B {\bf 642}, 197 (2006).

\bibitem{besiioo} M.~Ablikim {\it et al.} (BES Collaboration),
Phys. Lett. B {\bf 630}, 7 (2005).

\bibitem{pingrg} H.~Q.~Zhou, R.~G.~Ping and B.~S.~Zou,
Phys. Lett. B {\bf 611}, 123 (2005).

\bibitem{hsr}S.~J.~Brodsky, G.~P.~Lepage, Phys. Rev. D {\bf  24},
2848 (1981).
\bibitem{zhaoqiang} Xiao-Hai Liu and Qiang Zhao, Phys. Rev. D {\bf  81},
014017 (2010).

\bibitem{chendy} Dian-Yong Chen, Jun He, Xue-Qian Li and Xiang Liu, Phys. Rev. D {\bf  81},
074006 (2010).

\bibitem{bes3} M.~Ablikim {\it et al.} (BES Collaboration),
Nucl. Instrum. Meth.  A {\bf 614},  345 (2010).

\bibitem{geant4a} S. Agostinelli, {\it et al.} (GEANT4 Collaboration), Nucl. Instrum. Methods Phys. Res. A {\bf506}, 250 (2003).
\bibitem{geant4b} J. Allison, {\it et al.} IEEE Trans. Nucl. Sci. {\bf53}, 270 (2006).

\bibitem{psipdecays} M.~Ablikim {\it et al.} (BES Collaboration), Phys. Rev. D {\bf81}, 052005 (2010).
\bibitem{yellowbook} Kuang-Ta Chao and Yifang Wang, Inter. J. Mod. Phys. A {\bf 24}, Supplement 1 (2009).
\bibitem{evtgen}D. J. Lange, Nucl. Instrum. Methods Phys. Res., Sect. A {\bf462}, 152 (2001).
\bibitem{pdg2010}  K.~Nakamura {\it et al.} (Particle Data Group),
J. Phys. G: Nucl. Part. Phys. {\bf 37}, 075021 (2010).

\bibitem{roofit} W. Verkerke and D. P. Kirkby, arXiv:physics/0306116.
\bibitem{syserrors} M. Ablikim {\it et al.}, (BES Collaboration),  Phys.  Rev.  D {\bf
83}, 112005 (2011).
\end{thebibliography}
\end{document}